\documentclass[twocolumn,10pt]{revtex4} 
\usepackage{amsmath,amssymb,graphicx}
\begin{document}

\title{Quantum phase transition in a shallow one-dimensional optical lattice}

\author{Tomasz Sowi\'nski$^{(\alpha,\beta)}$} 
\affiliation{$^{(\alpha)}$Institute of Physics of the Polish Academy of Sciences \\ $^{(\beta)}$Center for Theoretical Physics of the Polish Academy of Sciences \\ Al. Lotnk\'ow 32/46, 02-668 Warsaw, Poland}

\begin{abstract}In this article the extended Bose-Hubbard model describing ultra-cold atoms confined in a shallow, one-dimensional optical lattice is introduced and studied by the exact diagonalization approach. All parameters of the model are related to the only relevant parameter controlled experimentally -- the depth of the optical potential. Changes in a shape of the insulating lobe in the phase diagram of the system are explored and the value of the critical tunneling for which the system undergoes the phase transition (from the insulating to the superfluid phase) is predicted. It is shown that the value of critical tunneling is substantially affected by the presence of the tunnelings to distant sites of the optical lattice. The results may have some importance in upcoming experiments on quantum quench through phase transition points.

\end{abstract}

\maketitle 

\section{Introduction}
The celebrated Hubbard model was introduced over fifty years ago to give an intuitive explanation of the transition from the superconducting to the insulating phase for not fully filled conducting bands \cite{Hubbard}. The model revealed that on the many-body level the quantum phase transitions are induced not only by a mobility of single particles but also by mutual interactions between them. Due to amazing experimental progress in controlling of ultra-cold gases in optical lattices, this old fashioned theoretical toy-model is undergoing a renaissance since it provides a realistic description of the real quantum systems \cite{Jaksch,Greiner,LewensteinBook}. Many different extensions of the standard Hubbard model came into a play due to a subtle analysis of interactions between particles. When interactions are strong enough they can induce inter-site and inter-band couplings and may lead to many interesting phenomena \cite{Luehmann,Mering,Bissbort,DuttaAndre,SowinskiMolecules,Pietraszewicz,SowinskiVibrating,LackiZakrzewski,Maik}. This path was intensively explored by many authors and recently it was summarized in a broad review \cite{Dutta}. Nevertheless, there are still open questions on the properties of ultra-cold gases confined in a shallow optical lattice. One suspects that in such a case the validity of the Hubbard-like Hamiltonian can be questionable and some other methods should be adopted. One possibility is to exploit some methods working directly in the configuration space of confined particles. This idea was adopted recently with so-called hybrid quantum Monte Carlo method and used to study the phase transition from the Mott-Insulator (MI) to the superfluid (SF) phase \cite{Pilati}.

In this article the old-fashioned Hubbard-like approach is applied to this problem. First, we start from the general Hamiltonian of the system and we rewrite it in the form of the exact multi-orbital Hubbard like model. Then we show that even for very shallow lattices, for the lowest commensurate filling of the lattice, $\rho=1$, in the vicinity of the phase transition the single-particle tunneling to the next-nearest-neighbor (NNN) sites has to be taken into account. Moreover, we show that NNN tunneling is a dominant correction to the standard Bose-Hubbard (BH) model and in the first approximation it is sufficient to neglect other corrections. In this way we obtain a quite simple extended Bose-Hubbard (BH) model and we study its properties in the vicinity of the phase transition with the exact diagonalization method.

The paper is organized as follows. In Sec. \ref{Sec:Model} the extended BH model with tunnelings to the NNN is introduced. In Sec. \ref{Sec:Results} the method of the exact diagonalization of the Hamiltonian is explained and an extrapolation of the results to the thermodynamic limit is clarified. In this way the phase diagram of the system for different lattice depths is calculated and the critical tunneling is localized. Finally, the conclusions are presented in Sec. \ref{Sec:Conclusions}. 

\section{The Model} \label{Sec:Model}

The starting point of the derivation of the model is a general Hamiltonian describing bosons of mass $m$ confined in an optical lattice potential and interacting mutually via short range delta-like potential. We assume that the dynamics of the system is completely frozen in directions perpendicular to the direction of the optical lattice. Experimentally it can be obtained by applying very strong harmonic confinements in these directions. Then, excitations to higher states of harmonic confinement are strongly suppressed and, in consequence, particles occupy only the ground-state of the harmonic oscillator. Effectively, the second-quantized Hamiltonian of the system in the direction of the optical lattice has a form:
\begin{align} \label{Ham0}
\hat{\cal H} &= \int\mathrm{d}x\, \hat\Psi^\dagger(x)H_0\hat\Psi(x)  + \frac{g}{2}\int\mathrm{d}x\,\hat\Psi^\dagger(x)\hat\Psi^\dagger(x)\hat\Psi(x)\hat\Psi(x),
\end{align} 
where $H_0=-\frac{1}{2m}\frac{\mathrm{d}^2}{\mathrm{d}x^2}+V\sin^2(kx)$ is a single-particle part of the Hamiltonian. Lattice intensity $V$ and the wavevector $k$ are controlled by the lasers forming an optical lattice. The effective one-dimensional contact interactions are controlled by the coupling constant $g$ \cite{Olshanii}. The bosonic field operator $\hat\Psi(x)$ annihilates a particle at point $x$ and fulfills the standard commutation relations for bosonic fields $[\hat\Psi(x),\hat\Psi^\dagger(x')]=\delta(x-x')$. For convenience the whole analysis will be performed in the natural units of the problem, i.e. we measure energies in recoil energies $E_R=\hbar^2k^2/2m$, lengths in $1/k$, etc. In this way the coupling constant $g$ is dimensionless and it measures an effective strength of the interaction between particles. 

The standard route to obtain the effective Hubbard-like model describing the system is to expand the field operator in the basis of the maximally localized Wannier functions ${\cal W}^{(\alpha)}_i(x)$. The Wannier functions are numbered with two discrete indices $\alpha$ and $i$ corresponding to the Bloch band and lattice site of the periodic potential, respectively. In the case studied, the functions can be found straightforwardly by solving the single-particle Schr\"odingier equation, which has a form of well-known Mathieu equation. The expansion of the field operator in the basis of Wannier functions has a form
\begin{equation} \label{Expansion}
\hat\Psi(x) = \sum_{i,\alpha} {\cal W}^{(\alpha)}_i(x) \hat a_{\alpha i},
\end{equation}
where $\hat a_{\alpha i}$ annihilates a boson in the single particle state described by the wave function ${\cal W}^{(\alpha)}_i(x)$. By putting expansion \eqref{Expansion} to the Hamiltonian \eqref{Ham0} one gets a general multi-band Hubbard-like Hamiltonian of the form
\begin{equation} \label{Ham1}
\hat{\cal H} = \sum_\alpha\sum_{ij} t^{(\alpha)}_{|i-j|}\hat{a}^\dagger_{\alpha i}\hat{a}_{\alpha j} + \sum_{ijkl}\sum_{\alpha \beta \gamma \delta} U^{(\alpha \beta \gamma \delta)}_{ijkl} \hat{a}^\dagger_{\alpha i}\hat{a}^\dagger_{\beta j}\hat{a}_{\gamma k}\hat{a}_{\delta l}.
\end{equation}
The parameters $t^{(\alpha)}_{n}$ and $U^{(\alpha \beta \gamma \delta)}_{ijkl}$ are appropriate matrix elements of the original Hamiltonian \eqref{Ham0}
\begin{subequations}
\begin{align}
t^{(\alpha)}_{l}&= \int\mathrm{d}x\,\overline{\cal W}^{(\alpha)}_i(x)\,H_0\,{\cal W}^{(\alpha)}_{i+l}(x), \\
U^{(\alpha \beta \gamma \delta)}_{ijkl} &= \frac{g}{2}\int\mathrm{d}x\,\overline{\cal W}^{(\alpha)}_i(x)\overline{\cal W}^{(\beta)}_j(x){\cal W}^{(\gamma)}_{k}(x){\cal W}^{(\delta)}_{l}(x).
\end{align}
\end{subequations}
The diagonal parameters $t^{(\alpha)}_{0}$ represent average single-particle energies in the states described by Wannier functions  ${\cal W}^{(\alpha)}_i(x)$. In contrast, the off-diagonal terms $t^{(\alpha)}_{n>0}$ are the tunneling amplitudes between lattice sites. Note, that due to the properties of Wannier functions the tunneling between lattice sites does not change the orbital $\alpha$. Since Wannier functions form a complete basis for the  single-particle problem the Hamiltonian \eqref{Ham1} is completely equivalent to the  Hamiltonian \eqref{Ham0}. The advantage is that, in this picture, we are dealing with localized states and therefore it is a quite clear procedure leading to the simplified models. 

The standard BH model is obtained in the so-called tight-binding limit, i.e. when all inter-band and inter-site interactions can be neglected and only the nearest-neighbor tunneling is important. In the experiments with ultra-cold bosons this scenario is achieved for deep enough lattices and weak enough interactions when all particles occupy the lowest Bloch band of periodic potential. In this limit only the on-site interaction in the lowest Bloch band $U^{0000}_{iiii}$ and the nearest-neighbor tunneling $t^{(0)}_1$ are important. For simplicity we will use standard notation $U$ and $t$, respectively. Properties of the BH model are well known and were studied in many different contexts and with the help of different numerical and analytical methods \cite{Fisher,Batrouni,Ejima}. 

Recently, due to the instantaneous experimental progress, many extensions of the standard BH model where introduced. Typically, extensions are needed whenever interactions between particles are enhanced and their long-range part starts to be important. In such scenarios the BH model is extended by an appropriate inter-site and inter-band terms. These path has opened very wide filed of research and has led to many interesting conclusions \cite{Dutta}.  

\begin{figure}
\includegraphics{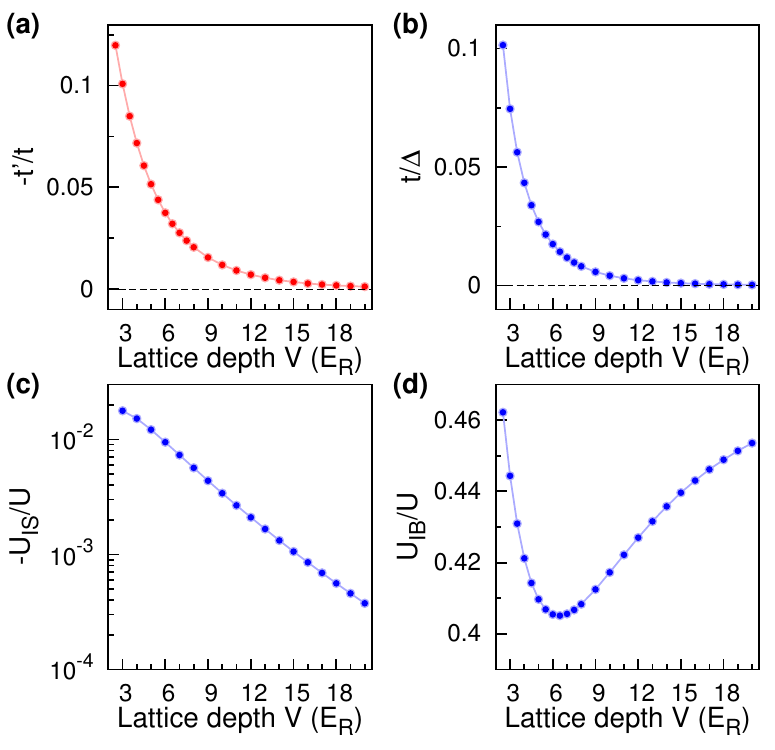}
\caption{(Color online) Most relevant parameters of the extended multi-orbital BH Hamiltonian \eqref{Ham1} as functions of the depth of the optical lattice. (a) Ratio of the NNN tunneling amplitude $t'$ to the tunneling amplitude $t$ (opposite sign included). For a very shallow lattices the NNN tunneling is of the order of 10\% of the tunneling $t$. (b) Ratio of the tunneling amplitude $t$ to the energy gap between bands $\Delta$. Even for shallow lattices the tunneling amplitude is much smaller than the energy gap. (c) Ratio of the interaction-induced inter-site tunneling $U_\mathtt{IS}$ to the local interaction parameter $U$. Due to the short-range character of the interactions all inter-site terms are negligible small. (d) Ratio of the inter-band coupling $U_\mathtt{IB}$ induced by interactions to the local interaction parameter $U$.  In all plots, the parameters are calculated numerically from the exact shape of the Wannier functions.  \label{fig1}}
\end{figure}

Here, we discuss a different regime of experimental parameters where interactions between particles are still very weak but the periodic potential of the optical lattice is very shallow. Our aim is to derive the most relevant corrections to the standard BH model and find their consequences in determining the position of the phase transition point from the MI to the SF phase. 

The depth of the optical lattice is directly related to the energetical spreading of the Bloch bands (spatial spreading of the localized Wannier functions) and is responsible for the narrowing of the energy gaps between them. Intuitively, for shallow lattices, one expects that the BH model needs to be extended by many different terms taking these simple observations into account. In general, such a model is quite complicated and hard to analyze. However, subtle analysis shows that in the vicinity of the quantum phase transition point (for filling $\rho=1$) many of new terms can be completely neglected and the only correction which is relevant comes from the tunnelings to the NNN $t_2^{(0)}$. In these particular case corrections that originate in the higher bands physics, as well as in the inter-site interactions, are less important and at the first approximation they can be omitted. 

To justify these non-obvious statements let us at first show that the influence of higher bands is negligble. 
The argumentation is based on the observation that the higher bands enter the game through interaction terms which lead to the promotion of particles from the ground to the higher bands. The most important term of this kind is related to the inter-band interaction $U_\mathtt{IB}=U_{iiii}^{(1100)}$, which is responsible for the inter-band transfer of two particles from the ground band $|\mathtt{g}\rangle=(\hat{a}_{0i}^\dagger)^2|\mathtt{vac}\rangle$ to the first excited band $|\mathtt{e}\rangle=(\hat{a}_{1i}^\dagger)^2|\mathtt{vac}\rangle$. In the subspace of two particle states in given lattice site the local Hamiltonian (truncated to the two lowest bands) has the form
\begin{equation}
\hat{\cal H}_{\mathtt{loc}} = 2\Delta |\mathtt{e}\rangle\langle\mathtt{e}| + 2U_\mathtt{IB}\left(|\mathtt{g}\rangle\langle\mathtt{e}|+|\mathtt{e}\rangle\langle\mathtt{g}|\right),
\end{equation}
where $\Delta$ is the energy gap between lattice bands. When the inter-band interaction $U_\mathtt{IB}$ is taken into account the true local ground state $|\mathtt{G}\rangle$ has some contribution from the excited state $|\mathtt{e}\rangle$ and the squared projection
\begin{equation}
|\langle\mathtt{e}|\mathtt{G}\rangle|^2 = \frac{x^2}{2x^2+2+2\sqrt{1+x^2}}, \qquad x=\frac{2U_\mathtt{IB}}{\Delta}
\end{equation} is a proper measure of the influence of higher bands. To estimate its value let us remind that, for one-dimensional case, the system undergoes the phase transition in the region where the ratio $t/U\gtrsim 0.25$. This ratio is independent on the lattice depth and for given lattice can be controlled by adjustment of the interaction coupling constant $g$. The energy gap $\Delta$ between the ground and the excited band of the periodic potential highly depends on the lattice depth. The ratio $t/\Delta$ as a function of the lattice depth $V$ is presented in Fig. \ref{fig1}b. As one can note, the tunneling $t$ is always much smaller than $\Delta$ and even for very shallow lattices it is not larger that $10\%$ of $\Delta$. At the same time the inter-band interaction $U_\mathtt{IB}$ is at least two times smaller than the on-site interaction $U$ (Fig. \ref{fig1}d). From these two facts and the following chain rule
\begin{equation}
\frac{U_\mathtt{IB}}{\Delta}=\frac{U_\mathtt{IB}}{U}\cdot\frac{U}{t}\cdot\frac{t}{\Delta}
\end{equation}
one finds that in the vicinity of the phase transition point ($U/t<4$) the interaction energy $U_\mathtt{IB}/\Delta<0.2$ for any lattice depth. Consequently, even for very shallow lattices, the squared projection $|\langle\mathtt{e}|\mathtt{G}\rangle|^2<4\%$. In addition, correction to the energy of the ground state, $|\delta E|/\Delta=1-\sqrt{1+x^2}<8\%$. 

This relatively small contribution from higher bands is caused mainly by an existence of a large energy gap $\Delta$. In other words, all processes that lead to the excitation of particles to the higher bands are always off-resonant and they are effectively suppressed by the conservation of energy. Note however, that this argumentation breaks down for larger fillings due to the enhancement of the interactions, i.e. the interaction parameters are multiplied by the numbers of particles occupying appropriate states. 

The situation is quite different for processes acting within the ground-band of the lattice. In this case, the inter-site processes (tunnelings or interactions) couple states with relatively equal energies and therefore they are not suppressed by the conservation of energy. They lead to the non-local correlations that have a crucial importance in the vicinity of the transition point from MI to SF phase. The most relevant process of this kind is obviously the single-particle tunneling to the neighboring site controlled by $t$. To find the leading correction to this term for shallow lattices we should compare the influence of the NNN tunneling $t'=t_2^{(0)}$ with the influence of the interaction induced tunneling $U_{\mathtt{IS}} =U_{iiij}^{(0000)}$ for $i=j\pm 1$ which transfers the particle to the neighboring site when the second particle is present nearby \cite{DuttaAndre}. As it is seen in Fig. \ref{fig1}c the inter-site amplitude $U_\mathtt{IS}$ is always much smaller than $U$. It is worth noting that induced tunneling $U_\mathtt{IS}$ is small directly due to the short-range character of mutual interactions. It is known that it can play a crucial role in the case of long-range interactions \cite{SowinskiMolecules}. 
It can be shown that, in the vicinity of the phase transition, $|U_\mathtt{IS}|/t$ is also essentially smaller than $|t'|/t$. Therefore, for shallow lattices, the most relevant correction comes from the NNN tunneling $t'$ and we will study the model with this correction only. Obviously, further improvement of the model (which is beyond the scope of this article) would require other terms. One of the most important would be the tunneling induced interaction $U_\mathtt{IS}$.  

At this moment it should be emphasized that, due to the properties of Wannier functions, the tunneling amplitude $t'$ has an opposite sign when compared with tunneling $t$. It means that the NNN tunneling effectively {\em decreases} the kinetic energy of the particles. As explained later, this fact has significant and counterintuitive consequences on the stability of the insulating phase. 

In this way we finally obtain the extended BH model for shallow lattice of the form:
\begin{align} \label{HamStudied}
\hat{\cal H} &= -t\sum_i \hat{a}_i^\dagger(\hat{a}_{i-1}+\hat{a}_{i+1}) -t' \sum_i \hat{a}_i^\dagger(\hat{a}_{i-2}+\hat{a}_{i+2}) \nonumber \\ &+ \frac{U}{2}\sum_i \hat{n}_i(\hat{n}_i-1).
\end{align}
Two tunneling amplitudes $t$ and $t'$ depend directly on the lattice depth $V$. In the tight-binding limit, when the NNN tunneling $t'$ is neglected, the model is effectively controlled by the one parameter $t/U$. The value of this parameter, along with the density of particles $\rho=N/L$, determines all properties of the ground-state of the system. From the experimental point of view this parameter can be controlled in two independent ways, i.e. by changing the depth of the optical lattice or by changing the interaction coupling parameter $g$. As long as $t'$ is excluded both experimental approaches are equally good from the model point of view. However, when the NNN tunneling is taken into account the Hamiltonian is controlled by two independent parameters $t/U$ and $t'/U$ and in the second experimental scenario (when only interaction coupling $g$ is changed) their ratio is fixed by the lattice depth, $t'/t=\mathrm{const}$. Therefore, one can still study properties of the system as a function of the normalized tunneling $t/U$ provided that the lattice depth $V$ is fixed. In the first experimental scenario when the lattices depth is changed, the ratio of both tunnelings $t'/t$ varies (see Fig. \ref{fig1}a). This simple observation should be always taken into account whenever a dynamical quench through the phase transition is considered. One should realize that two, seemingly equivalent, methods of controlling $t/U$ can lead to different results due to the neglected NNN tunnelings. In the following studies, we assume that the parameters of the Hamiltonian are controlled for the given lattice depth via changing interaction strength $g$.

At this point it is worth a reminder that the model is valid in the restricted range of parameters, i.e. for the first insulating phase ($\rho=1$) and in the vicinity of the phase transition point. Nevertheless, the model is sufficient to determine the value of critical tunneling for different lattice depths. 
 
\section{Method and Results} \label{Sec:Results}
\begin{figure}
\includegraphics{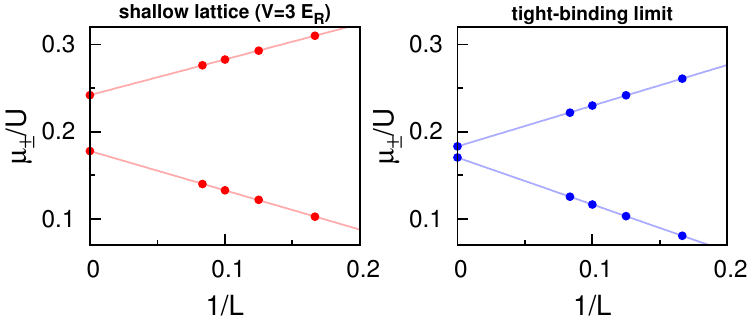}
\caption{(Color online) The upper ($\mu_+$) and lower ($\mu_-$) limits of the insulating phase as a function of the inverse of the system size $1/L$ for shallow (left panel) and deep optical lattice (right panel) at particular point $t/U=0.25$. The NNN tunneling $t'$ is calculated directly from the shape of the Wannier functions. The solid lines are linear fits to the numerical data points. Extrapolation to the infinite system size $1/L=0$ gives the energy gap $\Gamma$ of the insulating phase in the thermodynamic limit. Note that the energy gap $\Gamma$ is larger for shallow lattices. It means that, in contrary to the naive intuition, taking into account NNN tunnelings stabilizes the insulating phase.  \label{Fig2a}}
\end{figure}

As noted previously, here we assume that the lattice depth is fixed and the ratio $t/U$ is tuned by changing the local interaction term $U$. In this way the phase diagram for different lattice depths is obtained and the critical value of the tunneling as a function of the lattice depth is determined. For a very deep lattice the ratio $t'/t$ tends to 0 and one should expect to recover the limit of the standard BH model (tight-binding limit). 

The phase diagram of the model studied is obtained via a straightforward method based on the procedure of an  exact diagonalization of the Hamiltonian \cite{Elesin,Sowinski}. First, we fix the number of lattice sites $L$ and the number of particles $N$. We perform exact diagonalization of the Hamiltonian \eqref{HamStudied} in full many-body Fock space (i.e., each lattice site can be occupied with $0,\ldots,N$ particles) applying periodic boundary conditions. In this way we find the ground state of the system $|\mathtt{G}_{N,L}\rangle$ and its energy $E(N,L)$. Then, for commensurate filling $\rho=N/L=1$, we calculate upper and lower chemical potential $\mu_{\pm}$ defined as:
\begin{subequations}
\begin{align}
\mu_+(L)&=E(L+1,L)-E(L,L),  \\
\mu_-(L)&=E(L,L)-E(L-1,L).
\end{align}
\end{subequations}
These two quantities strongly depend on the lattice size $L$ and we are interested in their values in the thermodynamic limit of infinite lattice. Therefore, we diagonalize the Hamiltonian for different sizes of the lattice $L=4,8,10,12$ and we extrapolate data to the $L\rightarrow \infty$. The extrapolation procedure is based on the observation that, for large enough lattice sizes, the chemical potential scales linearly with the inverse of the lattice size $1/L$. In this way we obtain upper and lower bounds of the insulating lobe as a function of the normalized tunneling $t/U$. In this limit we also define the energy gap of the insulating phase $\Gamma=\mu_+-\mu_-$. In Fig. \ref{Fig2a} we present an example data points and the extrapolated values of $\mu$'s for $t/U=0.25$ in the lattice of infinite depth (tight-binding limit, $t'/t=0$) and with $V=3E_R$, respectively.

\begin{figure}
\includegraphics{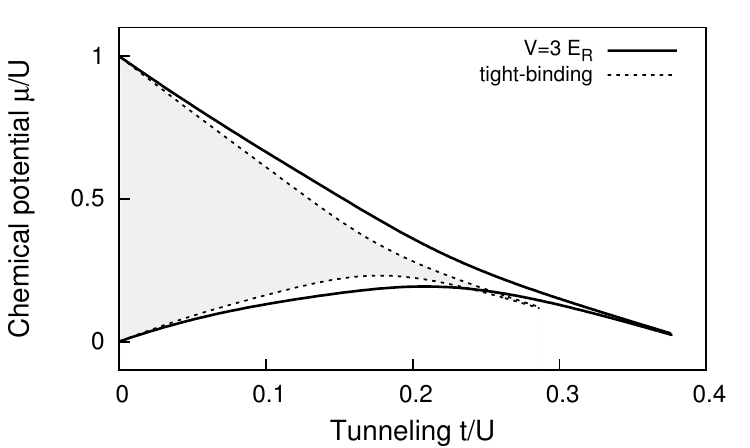}
\caption{Shape of the first insulating lobe in the phase diagram of the system in the case of a very shallow  lattice ($V=3E_R$) and in the tight-binding limit when NNN tunnelings can be completely neglected. Changing in the shape and the movement of the tip of the lobe is visible. The plot is obtained from the numerical data points by extrapolation to the thermodynamic limit ($L\rightarrow\infty$) as discussed in the text. \label{Fig2}}
\end{figure}

\begin{figure}
\includegraphics{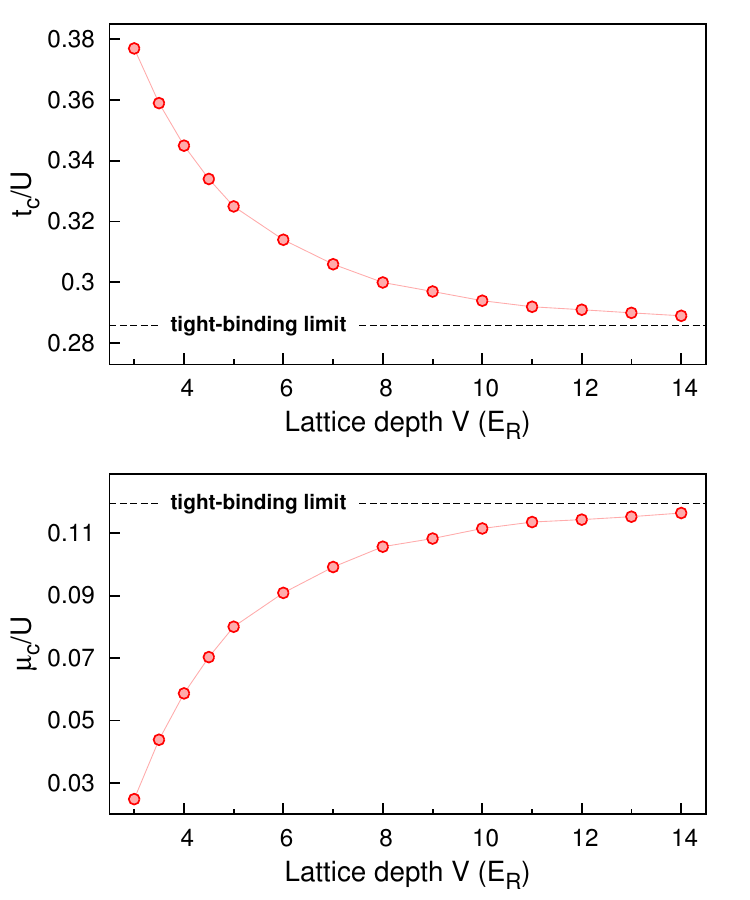}
\caption{(Color online) Position of the quantum phase transition point from the MI to the SF phase as a function of the depth of the optical lattice. For shallow lattices the critical tunneling $t_c$ is larger and the insulating phase is more stable (upper panel). In contrast to higher dimensions the critical value of the chemical potential decreases with the lattice depth (bottom panel). \label{Fig3}}
\end{figure}

In principle, the quantum phase transition point from the MI to the SF phase occurs in the system when the energy gap $\Gamma$ becomes equal to zero. Technically, this definition cannot be adopted directly to the numerical calculations due to the numerical uncertainty of $\Gamma$. Here, it is assumed that this uncertainty comes mainly from the extrapolating procedure to the thermodynamic limit. As a consequence, the position of the transition point $t_c$ is defined as the hopping amplitude at which the insulating gap becomes smaller than this uncertainty \cite{Sowinski}. In this way the phase diagram for a given lattice depth can be obtained and the position of the quantum phase transition can be estimated. In Fig. \ref{Fig2} the phase diagrams for two extreme lattice depths $V=3E_R$ and $V\rightarrow\infty$ (tight-binding limit) are presented. As one can note, the position of the tip of the insulating lobe is substantially affected by the presence of the tunnelings to the NNN. The position of the transition point estimated for different lattice depths is presented in Fig. \ref{Fig3}. For very strong optical potential the tight-binding limit is achieved. Moreover, it is interesting to note that, contrary to a naive intuition, for more shallow lattices (when the NNN tunnelings are enhanced) the insulating lobe is enlarged. It comes from the fact that the tunneling $t'$ has an opposite sign to the ordinary tunneling $t$ and some kind of destructive interference of both processes is present in the system. 
This fact was already noticed for an analogous model of a two-dimensional system on the basis of the quantum rotor model \cite{Kopec} and in the mean-field approximation \cite{Gao}. The position of the tip is also shifted in the direction of the chemical potential $\mu_c$. As it is seen in Fig. \ref{Fig3} for shallow lattices the critical point occurs for smaller chemical potential. This behavior is typical only for the one-dimensional case since for higher dimensions opposite result was predicted \cite{Kopec,Gao}.

\section{Conclusions} \label{Sec:Conclusions}

To conclude, in this paper the extended BH model describing the system of ultra-cold bosons in a shallow one-dimensional optical lattice was studied. It was shown that including additional single-particle tunnelings to the NNN is sufficient to describe the properties of the system in the vicinity of the quantum phase transition point for low densities (first insulating lobe). For a given depth of the optical lattice, the NNN tunneling amplitude was calculated from the exact shape of Wannier functions. To find the critical behavior of the system, exact diagonalization of the Hamiltonian was used. For shallow lattices the first insulating lobe is enlarged due to the opposite sign of the NNN tunneling and the critical value of the chemical potential $\mu_c$ is decreased. 

Results presented can shed some light on the problem of recent experiments and related theoretical works on the quench through the quantum phase transition point. Typically, such experiments are done by changing the optical lattice intensity and NNN tunnelings are neglected. In the first experiment with $^{87}$Rb atoms confined in a three-dimensional optical lattice \cite{Greiner} the quantum phase transition was located around $V\sim 13E_R$. In such a case NNN tunnelings seems to be very small (see Fig. \ref{fig1}a) and in the first approximation can be neglected. Nevertheless, it is worth noting that, whenever high precision measurement is done, these tunnelings introduce a quite big uncertainty of $2\%$ on the value expected from the tight-binding limit (Fig. \ref{Fig3}). This observation can have deep consequences for the problem of very precise quantum simulations performed in optical lattices \cite{Mark}. One should remember that the results obtained can be substantially affected by a presence of commonly neglected tunnelings to distant sites. As shown here, for high precision measurements, the quench process should be done by tuning of the interaction constant $g$ in a deep lattice with fixed intensity.

The author thanks M. Gajda for fruitful comments and suggestions. This research was supported by the (Polish) National Science Center Grant No. DEC-2012/04/A/ST2/00090.

\end{document}